# A software for learning Information Theory basics with emphasis on Entropy of Spanish

Fabio G. Guerrero, Member, IEEE and Lucio A. Pérez

Abstract—In this paper, a tutorial software to learn Information Theory basics in a practical way is reported. The software, called IT-tutor-UV, makes use of a modern existing Spanish corpus for the modeling of the source. Both the source and the channel coding are also included in this educational tool as part of the learning experience. Entropy values of the Spanish language obtained with the IT-tutor-UV are discussed and compared to others that were previously calculated under limited conditions.

F. G. Guerrero is with the Electrical Engineering School, Universidad del Valle, Cali, COLOMBIA (phone: 57.2.3392140 ext 109; fax: 57-2-3392361; e-mail: fguerrer@ univalle.edu.co).

L. A. Pérez is with IPTotal Software S.A. (www.iptotal.com), Cali, COLOMBIA (phone: 57.2.3309191 ext 105; fax: 57-2-6829757; e-mail: lucio.perez@ iptotal.com)

## I. INTRODUCTION

THE teaching of information theory basic concepts is usually based on mathematical derivations. This requires the students to have a considerable capacity of abstraction in order to understand thoroughly. For example, concepts such as the entropy rate of a language are often presented just as theoretical concepts.

A more practical way to teach the information theory basic concepts was tested in the graduate course Information Theory (class code 710599) at the Electrical and Electronic Engineering Department at Universidad del Valle (Cali, Colombia). It is based on the use of the educational software IT-tutor-UV, which was developed at this department too.

The IT-tutor-UV was developed using Matlab 7.0. It runs in the Windows operating system. The Spanish word frequency database from Alameda and Cuetos [1] was used for the generation of Spanish text in the IT-tutor-UV. This database contains the frequency of the 81323 more common Spanish words. It was obtained from a corpus of 1950375 words which appeared in different representative sources of the Spanish literature such as newspapers, transcriptions, etc.

Also, with the IT-tutor-UV, a more updated approximation for the different values of the entropy of Spanish such as the trigram entropy and the word entropy was obtained.

The section II of this paper is a brief description of the IT-tutor-UV's main components. In Section III, updated values of the entropy of Spanish, which were obtained using the IT-tutor UV, are presented. Finally, in section IV, conclusions and lines for future work are presented.

II.     IT-TUTOR-UV STRUCTURE

The IT-tutor-UV has four logical modules: the discrete source, the source coding, the channel coding, and the channel as shown in Figure 1.

Fig. 1. IT-tutor-UV block diagram.

The discrete source is modeled as a stochastic process using a first-order word approximation (i.e., words that have the right frequency in Spanish and that are considered statistically independent). The user can specify the number of words of the generated text sequence. First, this module counts the letter frequency, the digram frequency, the trigram frequency, or the word frequency according to the selection made by the user through the graphical user interface. Then, the observed entropy is calculated using the classical entropy equation (1) for each case.

$$H = \sum_i p_i \log p_i \qquad (1)$$

The discrete source module can also generate an Excel file with the frequency of the symbols found in the generated text. It is also possible to display and save the word sequence as a text file for later use. Before this is done, the database file should be loaded using the Windows™ ODBC Data Source Administrator. Fig. 2 shows a snapshot of the IT-tutor-UV graphical user interface.

Fig. 2. IT-tutor-UV graphical user interface.

Due to the long time that it can take to process a long sample when running the discrete source, the source coding, and the channel coding modules at the same time, the IT-tutor-UV has a box switch to use the discrete source module alone. This is useful to quickly observe the letter, digram, or trigram entropy. Fig. 3 shows a fifty word text sample produced by the IT-tutor-UV.

Fig. 3. Fifty word text sample.

The text sample in Fig. 3 conveys no meaning. The reason is that this text was produced using a first-order word approximation which lacks some language rules. These rules give a dependency between words that are close to each other. Then, if this dependency were included in the discrete source model, the text produced by it could eventually convey some meaning.

The source coding module supports both the Huffman coding method as described in [2] and the Arithmetic coding method [5], allowing the user to select the symbols for the coding process between letters, digrams, trigrams, or words.

In the channel coding module, the Reed-Solomon (RS) and convolutional coding techniques are included. With the RS coding, users are able to try different values for the RS word length $k$ (e.g., 5, 9, or 13), so they can verify the error correction capacity of the block coding when different amounts of redundancy are used. With the convolutional coding, two simple configurations of codes, which are shown in Fig. 4 and Fig. 5, can be chosen by the users.

Students can observe the effect of the redundancy on error correction capacity of convolutional codes and can make some comparisons with block coding not only on the basis of error correction capacity but also on the coding time.

Fig. 4. Convolutional code, Rate= ¼, $K$=3.

Fig. 5. Convolutional code, Rate = 2/3, $K$=[4,3].

For educational purposes, the channel is modeled as a binary symmetric channel with error probability $p$. The value of p can be entered using the graphical user interface. Once the message has been coded and sent through this channel, the IT-tutor-UV can display the bits transmitted, the error bits introduced by the channel, the BER (Bit Error

Rate), and the final error bits remaining after the message has been decoded. These values are used to calculate the correction capacity of the codes, which is displayed as a percentage of correction. Finally, a symbol-to-symbol comparison between the original source file and the decoded file is made.

III.     USING THE IT-TUTOR-UV

A.     Letters, digrams, and trigrams

Table I shows some values of the entropy for letters, digrams, and trigrams obtained using the IT-tutor-UV. The time necessary to calculate these results depends on the number of letters forming the symbols: the larger the number of letters, the larger the time the computer takes to finish the calculations. For instance, the calculation for the trigrams on Table I took 7.25 hours on a laptop with 1 GB of RAM memory and a 64-bit Processor at 2.2 GHz. The text sample's length in words was long enough to obtain the convergent values of the entropy for each of the cases in Table I.

TABLE I
LETTER, DIGRAM, AND TRIGRAM ENTROPY OF SPANISH

The IT-tutor-UV counts the accented vowels (á, é, í, ó, ú) as different symbols because these vowels give different meanings to the words of the Spanish language. The IT-tutor-UV also differentiates the "ñ" letter from the "n" letter. Then, the total of symbols is

33. These differentiations provide more information of the symbol frequency than the one provided in other tables which usually consider an alphabet of only 26 letters [3].

As can be seen on Table I, the single letter entropy obtained using the IT-tutor-UV was 4.15 bits/letter. This entropy value is slightly greater than the 4.0 bits/letter entropy obtained when considering only 26 letters. For digrams, the obtained entropy was 3.66 bits/letter. For trigrams, the obtained entropy was 3.38 bits/letter. The computation of this last entropy value was based on 3821 trigrams.

Fig. 6 shows a snapshot of the Excel trigram file produced by the IT-tutor-UV which was obtained from a 50000 word text sample. The 27 most frequent trigrams can be seen in Fig. 6. In this text sample, there were 3821 trigrams, and the total sum of the trigram frequencies was 138915.

Fig. 6. Trigram frequency example.

B. Letter Entropy per Word

Although the source model is working with a database of 81323 words, the calculated probability of the least frequent words such as those that appear once in the 1.95 million word corpus might not be precise enough. Therefore, finding the letter entropy per word of Spanish, $F_W$, by directly measuring the word frequencies from a very large text sample produced by the IT-tutor-UV would lead to a slightly inaccurate result.

To overcome this limitation, the Barnard's method [4] was used. This method estimates the letter entropy per word of a language using a limited word frequency list. It needs the log-log probability curve of the language. Fig. 7 shows this curve for the Spanish language, which was calculated using the word frequency database [1] employed by the IT-tutor-UV.

Fig. 7. Word rank versus word probability.

Using Barnard's method, the letter entropy per word can be calculated using (2).

$$F_W = \frac{-1.44269}{\alpha}\left[\ln k - k\left(\sum_{1}^{J}\frac{\ln n}{n} + \left\{\frac{(\ln M)^2}{2} - \frac{(\ln(J+0.5))^2}{2}\right\}\right)\right] \quad (2)$$

In (2), α is the average number of letters per word, $J$ is the number of words in the dictionary (i.e., 81323 words), $k$ is the language constant, and $M$ is the estimated number of words that would lead to a total probability of 1 as in (5). The values of α, and $k$ are given in (3) and (4) respectively.

$$\alpha = \sum_{1}^{81323} L_i P_i = 4.6978 \quad (3)$$

where $L_i$ is the length of the i-th word with probability $P_i$.

The language constant $k$ is a value such that the probability $P_n$ of the n-th word can be approximated as

$$k \approx \frac{p_n}{n} \quad (4)$$

and

$$\sum_{1}^{M} p_n = 1 \quad (5)$$

From data from Figure 7 a value for *k* of 0.0817 was obtained using the word range 1 to 1000. This range was chosen because this is the range in which *k* is most stable according to (4).

Following Barnard's method to find $F_W$, the value of ln *M* is solved using (6).

$$\ln(M) = \frac{1}{k} - \sum_{1}^{J} \frac{1}{n} + \ln(J + 0.5) \quad (6)$$

Then,

$$\ln(M) = \frac{1}{0.0817} - \sum_{1}^{81323} \frac{1}{n} + \ln(81323 + 0.5) \approx 11.6627$$

Finally, solving (2), $F_W$ is equal to

$$F_W = \frac{-1.44269}{4.6978}\left[\ln 0.0817 - 0.0817\left(\sum_{1}^{J} \frac{\ln n}{n} + \left\{\frac{(11.6627)^2}{2} - \frac{(\ln(81323.5))^2}{2}\right\}\right)\right]$$

$F_W \approx 2.4737$ bits/letter.

The obtained letter entropy per word of Spanish ($F_W$) is greater than the 1.97 bits/letter reported in [4], which is the only previous estimation of $F_W$ known by the authors. The difference is due mainly to the use of a much greater word frequency database in the IT-tutor-UV. Then, the entropy $F_W$ obtained here is a more precise value.

C.  Entropy Rate

In order to introduce the entropy rate concept [5] to students, the entropy per word of the IT-tutor-UV's source is calculated using (1) as follows

$$\sum_{1}^{81323} p_i \log_2 p_i = 10.4281.$$

Then, the entropy per letter is $10.4281/\alpha = 2.2198$ bits/letter. This value is the entropy rate, $H_W$, for this source too because the symbols it generates (i.e., words) are independent.

In practice, source coding is done by compressors. Then, the instructor could challenge students to find the compressor with the best compression ratio. For example, from all the compression programs the authors tested on a 10 million word file without spaces created by IT-tutor-UV, the best result was obtained using PPMonstr by Dmitry Shkarin [6]. This compressor was able to compress to 2.402 bits/letter, a value quite close to the 2.2198 bits/letter limit.

D.   A Source Coding and Channel Coding Example

When using the Huffman coding, the IT-tutor-UV generates the Huffman dictionary displaying the longest codes, the shortest codes, the average code length, and the compression rate which is calculated by comparing the sizes of the original and coded file. Fig. 8 shows a snapshot of a Huffman dictionary in the Excel file generated by the IT-tutor-UV program using words as source symbols.

Fig. 8.  Huffman dictionary file generated by IT-tutor-UV

The following example shows the results for a short text file test:

File size: 10000 words

Source coding: Huffman method

Source symbols: words

Channel Coding: Convolutional (rate=2/3, $K$= [4, 3])

BSC Error Probability: 0.005

Source Entropy: 8.009 bits/symbol

Average Code Length: 8.022 bits/symbol

Shortest code: 1011

Largest code: 00100000001

Original File Size: 5749 bytes

Encoded File Size: 1003 bytes

Compression rate: 82.5535%

Number of transmitted bits: 12033

Error Bits: 72

Error Bits after correction: 1

BER: 0.00012466

Correction rate: 98.6111%

Source file ≠ Destination File

From these values, it can be seen how Huffman coding closely approaches the source entropy when the coder's dictionary (i.e., the matching between code symbols and source symbols that the receiver needs to recover the text into its original form) is not considered.

The Arithmetic coding algorithm assigns a label to the whole sequence of symbols in the message. This label is generated according to the frequency of each symbol; therefore, there is only a single code for the entire message, and a code dictionary is not necessary.

Using the IT-tutor-UV, students can verify how close the arithmetic coding is to the entropy limit and what its advantages over Huffman coding are.

The IT-tutor-UV's executable code can be downloaded at [7] and used under a GPL license.

IV.  CONCLUSION

The IT-tutor-UV is an educational software for making the experience of learning some of the highly abstract concepts of information theory such as discrete source generation, information entropy, and channel capacity, a more practical and enjoyable experience.

Using the IT-tutor-UV, updated entropy calculations of the Spanish language such as the word entropy and trigram entropy were obtained.

Also, the IT-tutor-UV is very useful for obtaining complete frequency lists of letters, digrams, and trigrams of Spanish. These lists can be useful for different goals such as the study of classic decryption systems based on frequency.

The future work for this software is the introduction of a more complex discrete source module which is able to consider the grammar of Spanish in a rigorous way. Also, the addition of classic cryptography schemes to the IT-tutor-UV would be useful to learn about the security aspects of Information Theory.

REFERENCES

[1]     Cuetos F., and Alameda J. R. (1995). Diccionario de las unidades Lingüísticas del Castellano: Volumen I: Orden Alfabético / Volumen II: Orden por Frecuencias. [Online]. Available: http://www.uhu.es/jose.alameda.

[2]     Huffman A. David, "A Method for the Construction of Minimum-Redundancy Codes".  Proc.  IRE,  vol 40, pp 1098-1101, Sept. 1952.

[3]     Ferrero J.  "La lengua española y los medios audiovisuales". Congreso de la Lengua Española. Sevilla, 1992 [Online]. Available: http://cvc.cervantes.es/obref/congresos/sevilla/comunicacion/mesaredon_ferrero.htm.

[4]     G. A. Barnard, III, "Statistical calculation of word entropies for four western languages," IEEE Trans. Information Theory, vol. 1, pp. 49–53, Mar. 1955.

[5]     Wells R. B., Applied Coding and Information Theory for Engineers.   Upper Saddle River, New Jersey: Prentice Hall, 2002, ch. 1.

[6]     Shkarin D. (April 2002). Monstrous PPMII compressor based on PPMd var.I. [Online]. Available: http://compression.ru/ds/.

[7]     Guerrero F. and Perez L. "IT-tutor-UV instructional software" (March 2006). [Online]. Available: http://sistel-uv.univalle.edu.co/it-tutor-uv.html

Fabio G. Guerrero (M'92)  Born in Cali (Colombia). Graduated with a B.S. in Telecommunications Engineering from Universidad del Cauca, 1992, and a MSc. in Real -Time Electronic Systems from Bradford University, UK, 1995. Currently, he is with Universidad del Valle (Cali-Colombia) as assistant telecommunications lecturer. His main fields of interest are digital communications and telecommunication systems modeling.

Lucio A. Pérez Born in Pasto (Colombia). Graduated with a B.S. in Electronic Engineering from Universidad del Valle, 1997. Currently, he works as CTO at IPTotal Software S.A. (www.iptotal.com), a telecommunications oriented software development company.  His main fields of interest are software development and deployment for next generation networks focused on mediation, assurance and provisioning systems.

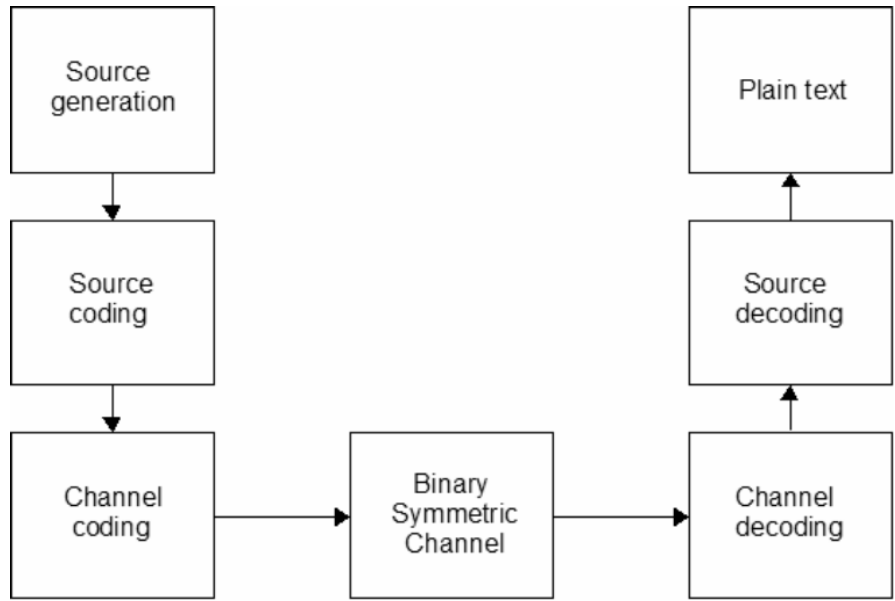

Fig. 1. IT-tutor-UV block diagram.

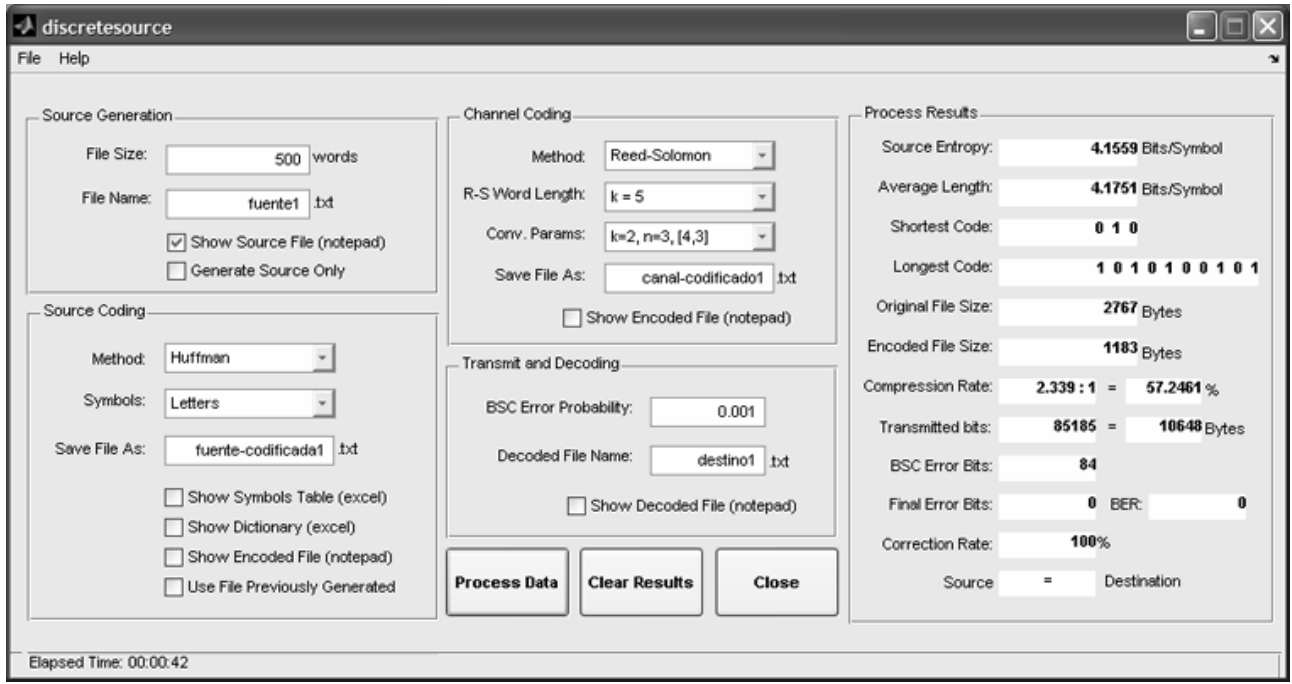

Fig. 2. IT-tutor-UV graphical user interface.

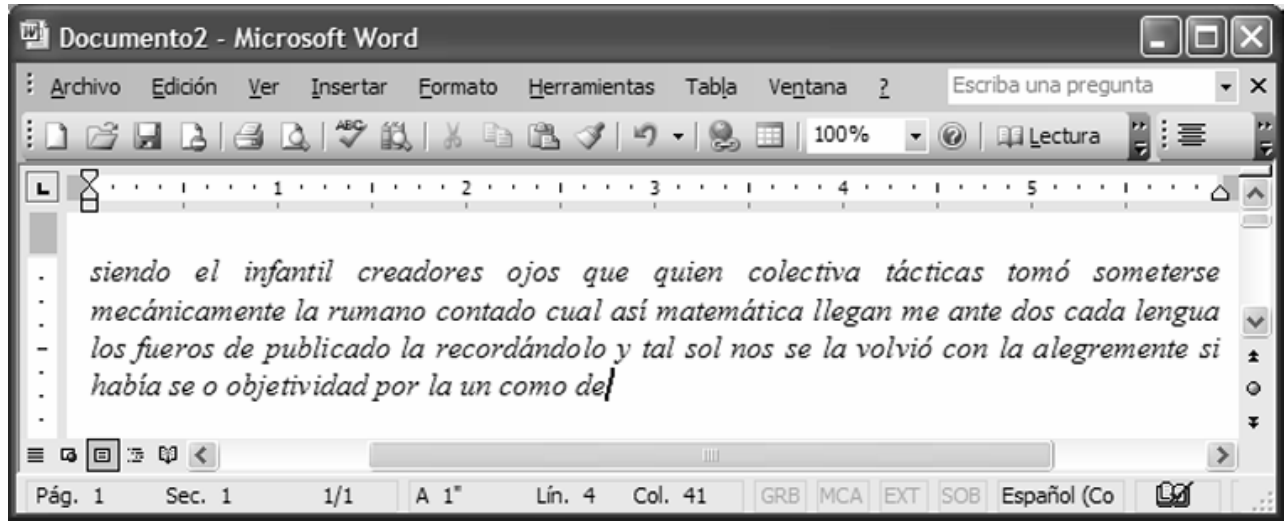

Fig. 3. Fifty word text sample.

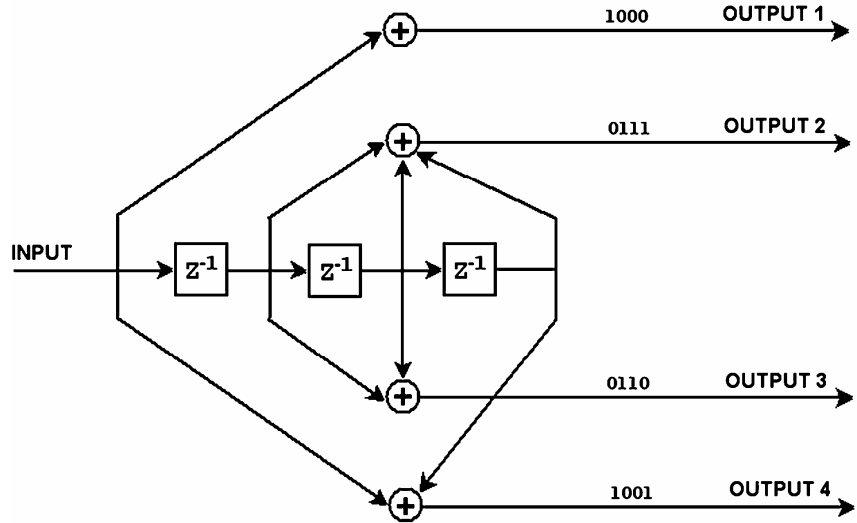

Fig. 4. Convolutional code, Rate= ¼, K=3.

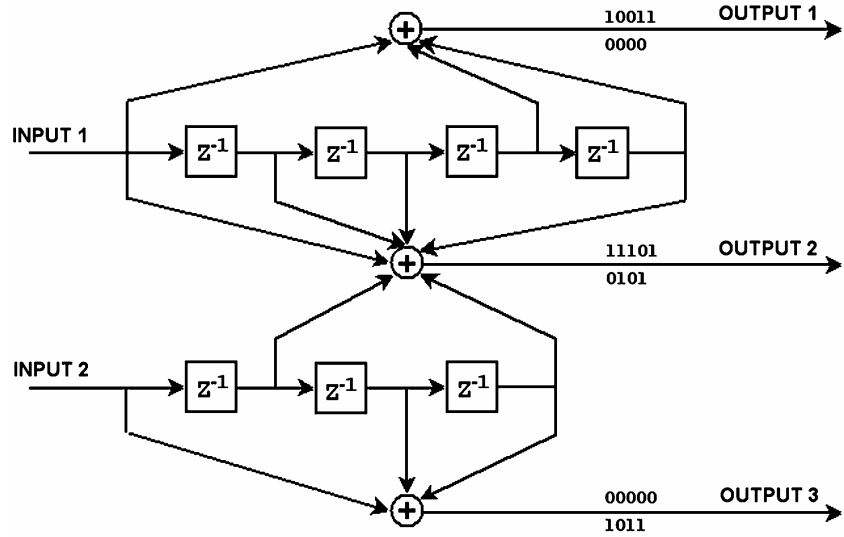

Fig. 5.  Convolutional code, Rate = 2/3, K=[4,3].

| | A | B | C |
|---|---|---|---|
| 1 | Símbolo | Frecuencia | P(i) |
| 2 | | | |
| 3 | que | 2138 | 0.015391 |
| 4 | ent | 1505 | 0.010834 |
| 5 | con | 1201 | 0.008646 |
| 6 | nte | 1188 | 0.008552 |
| 7 | los | 1010 | 0.007271 |
| 8 | est | 896 | 0.00645 |
| 9 | ado | 830 | 0.005975 |
| 10 | las | 738 | 0.005313 |
| 11 | ien | 721 | 0.00519 |
| 12 | por | 713 | 0.005133 |
| 13 | era | 681 | 0.004902 |
| 14 | par | 665 | 0.004787 |
| 15 | tra | 655 | 0.004715 |
| 16 | sta | 640 | 0.004607 |
| 17 | res | 639 | 0.0046 |
| 18 | men | 637 | 0.004586 |
| 19 | ión | 625 | 0.004499 |
| 20 | ndo | 618 | 0.004449 |
| 21 | ant | 593 | 0.004269 |
| 22 | una | 591 | 0.004254 |
| 23 | com | 574 | 0.004132 |
| 24 | per | 572 | 0.004118 |
| 25 | aci | 561 | 0.004038 |
| 26 | aba | 552 | 0.003974 |
| 27 | ida | 519 | 0.003736 |
| 28 | nto | 517 | 0.003722 |
| 29 | ció | 516 | 0.003715 |
| 30 | del | 515 | 0.003707 |

Fig. 6. Trigram frequency example.

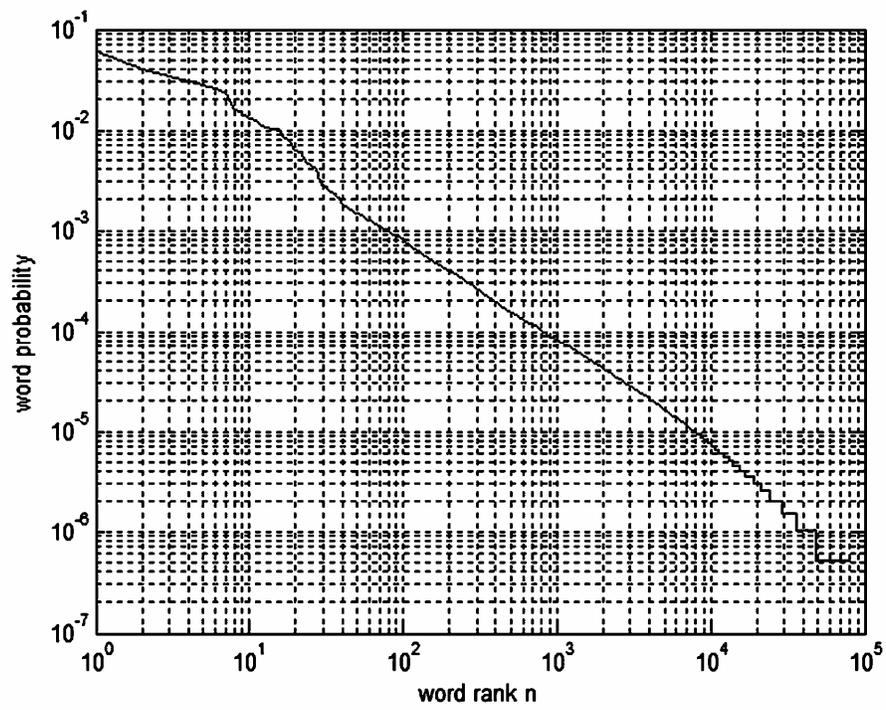

Fig. 7. Word rank versus word probability.

| Symbol | B | C | D | E | F | G | H | I | J | K | L | M | N | O | P | Q |
|---|---|---|---|---|---|---|---|---|---|---|---|---|---|---|---|---|
| | | | | | | | Code | | | | | | | | | |
| de | 1 | 1 | 1 | | | | | | | | | | | | | |
| para | 0 | 0 | 1 | 1 | 1 | 0 | 0 | | | | | | | | | |
| se | 0 | 0 | 0 | 0 | 0 | 0 | | | | | | | | | | |
| en | 0 | 0 | 1 | 0 | 1 | | | | | | | | | | | |
| sentido | 0 | 0 | 1 | 1 | 1 | 0 | 0 | 1 | 1 | 1 | 0 | | | | | |
| cada | 1 | 0 | 1 | 1 | 0 | 1 | 0 | 0 | 0 | | | | | | | |
| objeto | 0 | 0 | 0 | 1 | 0 | 0 | 0 | 0 | 0 | 0 | 1 | | | | | |
| hechos | 0 | 0 | 0 | 1 | 0 | 1 | 1 | 0 | 0 | 0 | 0 | 0 | 0 | 0 | | |
| un | 0 | 0 | 1 | 0 | 0 | 0 | | | | | | | | | | |
| con | 0 | 1 | 1 | 0 | 0 | 0 | | | | | | | | | | |
| y | 1 | 1 | 0 | 0 | | | | | | | | | | | | |
| piel | 0 | 0 | 0 | 1 | 0 | 1 | 1 | 1 | 1 | 1 | 0 | 0 | | | | |
| también | 0 | 0 | 1 | 1 | 0 | 0 | 1 | 1 | 0 | | | | | | | |
| hay | 1 | 0 | 0 | 0 | 1 | 0 | 0 | 0 | 1 | | | | | | | |
| el | 0 | 0 | 0 | 0 | 1 | | | | | | | | | | | |
| recuerdo | 0 | 0 | 1 | 1 | 0 | 0 | 0 | 1 | 0 | 1 | 0 | 1 | | | | |
| momento | 1 | 0 | 0 | 0 | 1 | 1 | 1 | 0 | 0 | 1 | | | | | | |
| me | 1 | 0 | 0 | 1 | 0 | 0 | 0 | | | | | | | | | |
| la | 0 | 1 | 1 | 1 | | | | | | | | | | | | |
| a | 0 | 1 | 0 | 0 | 1 | | | | | | | | | | | |
| más | 1 | 0 | 0 | 0 | 0 | 1 | 0 | | | | | | | | | |
| al | 0 | 0 | 1 | 0 | 0 | 1 | 0 | | | | | | | | | |
| son | 0 | 0 | 1 | 1 | 0 | 0 | 1 | 0 | 1 | | | | | | | |
| yo | 1 | 0 | 0 | 1 | 1 | 1 | 0 | 0 | | | | | | | | |
| cuando | 1 | 0 | 0 | 0 | 1 | 0 | 1 | 0 | | | | | | | | |
| ni | 0 | 0 | 0 | 1 | 0 | 1 | 0 | 1 | 1 | | | | | | | |
| las | 0 | 1 | 0 | 1 | 1 | 0 | | | | | | | | | | |
| su | 1 | 0 | 1 | 1 | 0 | 0 | | | | | | | | | | |
| ella | 0 | 0 | 0 | 0 | 0 | 1 | 0 | 0 | 0 | | | | | | | |
| están | 0 | 0 | 0 | 1 | 0 | 1 | 0 | 0 | 0 | 1 | 1 | | | | | |
| días | 1 | 1 | 0 | 1 | 1 | 0 | 1 | 0 | 1 | 0 | | | | | | |
| niño | 0 | 0 | 1 | 0 | 0 | 1 | 1 | 0 | 1 | 1 | 1 | | | | | |
| misma | 0 | 0 | 1 | 0 | 0 | 1 | 1 | 0 | 1 | 1 | 0 | | | | | |
| vanidad | 0 | 0 | 0 | 1 | 1 | 0 | 0 | 0 | 1 | 0 | 1 | 1 | 1 | 0 | 0 | 1 |

Fig. 8. Huffman dictionary file generated by IT-tutor-UV

| Symbol | Sample Length (words) | $H_{symbol}$ | $H_{letter}$ |
|---|---|---|---|
| Letters | 200000 | 4.1484 | 4.1484 |
| digram | 100000 | 7.3223 | 3.6612 |
| trigram | 50000 | 10.1432 | 3.3811 |

TABLE I

LETTER, DIGRAM, AND TRIGRAM ENTROPY OF SPANISH